\documentclass[12pt]{iopart}
\usepackage{graphicx,latexsym,stmaryrd,iopams}
\newcommand{\be}{\begin{eqnarray}}
\newcommand{\ee}{\end{eqnarray}}
\def\refeq#1{(\ref{#1})}

\def\la{\lambda}

\def\g{\gamma}
\def\al{\alpha}

\def\l{\left}
\def\r{\right}
\def\te{\mbox{e}}

\begin{document}

\title{Collective dispersion relations for the $1$D interacting two-component Bose and Fermi gases}

\author{M.T. Batchelor, M. Bortz, X.W. Guan and  N. Oelkers}
\address{\dag\
{\small Department of Theoretical Physics, RSPSE, and}\\
{\small Department of Mathematics, MSI}\\
{\small The Australian National University, Canberra ACT 0200, Australia}}

\date{\today}

\begin{abstract}
 \noindent
We investigate the elementary excitations of charge and spin degrees for the $1$D interacting 
two-component Bose and Fermi gases by means of the discrete Bethe ansatz equations.  
Analytic results in the limiting cases of strong and weak interactions are derived, where the Bosons are treated in the repulsive and the fermions in the strongly attractive regime.
We confirm and complement results  obtained previously from the Bethe ansatz equations in the thermodynamic limit. 
\end{abstract}



\maketitle 

\section{Introduction}

Recent achievements in trapping quantum gases of ultra-cold atoms have 
opened up many exciting possibilities for the experimental
investigation of quantum effects in low-dimensional many-body systems.
In particular, the experimental realization of the 
Tonks-Girardeau gas \cite{Exp-B1,Exp-B2,Exp-B3,Exp-B4,Exp-B5}, 
collective spin wave excitations in the spinor Bose gas \cite{Exp-SB1,Exp-SB2,Exp-SB3}, 
the quasi $1$D Fermi gas \cite{Fermi-1D1,new-Fermi} and the 
observation of atom matter waves \cite{M-W} 
have shed light on understanding the quantum nature of $1$D many-body systems.
There has been a corresponding revival of interest in the exactly solved models of interacting 
Bose and Fermi gases, such as the Lieb-Liniger Bose gas \cite{LL}, the Gaudin-Yang Fermi gas \cite{GY},
the $1$D Hubbard model \cite{Hubb,Amico} and the BCS pairing model \cite{BCS}, due to the fact that their
ground state and low lying excitations are accessible from exact Bethe ansatz solutions.

Quantum gases with multi-spin (pseudo-spin) states exhibit even richer quantum 
effects than their single component counterparts. 
The update facilities allow one to create multi-component Bose gases in low dimensions, 
e.g., two atomic hyperfine states can make up a pseudo-spin doublet \cite{Exp-SB1,Exp-SB2,Exp-SB3}. 
The observation of collective dynamics of spin waves in the spinor Bose gas has
stimulated theoretical attention on the $1$D integrable
spinor Bose gas \cite{The-SB1,The-SB2,Ieda-1,The-SB3}. 
In this paper, we present a
systematic way to derive dispersion relations of low-lying elementary excitations of the
spinor Bose and Fermi gases via the discrete Bethe ansatz equations
in the strong and weak coupling limits.
This approach has some distinct advantages compared to the usual way of deriving
the low-lying excitation spectrum: 
(i) it leads to explicit forms of the quasimomenta which are helpful in understanding the collective
behaviour of low-lying excitations; 
(ii) it avoids the counting of quantum numbers involved in the integral equation formulation 
of the Bethe equations, and
(iii) physical quantities such as the charge velocity, spin velocity, effective mass, Fermi momentum,
 finite-size corrections, ground state energy and excitation energies
 readily follow in the weak- and strong-coupling limits. Note that especially in weak coupling limit, those quantites are  difficult to deal with in the framework of the Bethe ansatz equations in the thermodynamic limit (due to singular integral kernels in the linear integral equations).

To describe elementary excitations in terms of their dispersion relations, consider the general case of $\mathcal M$ branches (or rapidities) of excitations labeled by 
$\mu=1,\ldots \mathcal , M$. 
For example, $\mathcal M=1$ for the spinless Bose gas (charge excitations only),
$\mathcal M=2$ for the spinor Bose and spin-$\frac12$ Fermi gas (spin and charge excitations). 
Generally, the relativistic dispersion relation for each one of these excitations is written as
\begin{eqnarray}
\varepsilon_\mu(p)=\sqrt{\Delta_\mu^2+v_\mu^2 p^2}\label{disp}
\end{eqnarray}
where $\Delta_\mu$ is the excitation gap and $v_\mu$ is the velocity. 
For vanishing gap $\Delta_\mu=0$, the spectrum is linear, such that
\be
v_\mu&=&\partial_p \varepsilon_\mu(p)\label{vmu}.
\ee 
In this case, the underlying effective field theory is conformally invariant. 
Examples are the charge channel for spinless bosons with repulsive interaction \cite{Haldane}, the charge channel for fermions and the spin channel for repulsive fermions. 
In the other extreme $\Delta_\mu\gg 1$, one ends up with
\begin{eqnarray}
\varepsilon_\mu(p)=\Delta_\mu+\frac{v_\mu^2 p^2}{2 \Delta_\mu}
:=\Delta_\mu+\frac{p^2}{2 m_\mu^*}\label{disp-2}
\end{eqnarray}
which is the classical dispersion of free particles with an effective mass $m_\mu^*$. 
It is known that this scenario is realized for spin excitations in the spinor Bose gas 
with repulsive interaction \cite{SW-1,SW-2}. This case has been analyzed recently \cite{The-SB2,The-SB3}.
We will consider this model with our method in Section 2, thereby confirming quantitatively the following results.
In the weak coupling limit, the effective mass of the spin wave is almost the same as the mass of one
boson, while for strong coupling, it tends to the total mass of all  bosons. 
On the other hand, charge excitations in this system are still gapless. 

In Section 3, we will turn our attention to the intermediate regime with finite $\Delta_\mu$, which is encountered for spin 
excitations in the attractive spin-$\frac12$ Fermi gas. 
In this system, the gap is an increasing function of $c^2$. 
The charge excitations again remain gapless.  
For the sake of comparison, we briefly recall the calculation of the
velocities for the attractive spin-$\frac12$ Fermi gas in the thermodynamic limit at the end of Section 3.  
We note that the method we propose to calculate $\varepsilon(p)$ in (\ref{disp}) can also be  
adapted to other models such as the $1$D integrable Hubbard model \cite{Hubb,Amico} and 
the mixed Bose-Fermi model \cite{L-Y,mix-1,mix-2,mix-3}.

\section{The spinor Bose gas} 
Before turning to the spinor Bose gas, we give an outline of the method we apply.

\subsection{Outline of the method}
\label{sec21}

In the framework of the Bethe ansatz solution, the energy eigenvalues $E$ and momenta $p$ of the two-component Bose and Fermi gases are given in terms of two
sets of Bethe roots, $\l\{k_j\r\}_{j=1,\ldots,N}$ and  $\l\{\la_\al\r\}_{\al=1,\ldots,M}$, 
corresponding to charge and spin degrees of freedom. Namely,
\be
E&=&\sum_{j=1}^N k_j^2\label{en}\\
p&=& \sum_{j=1}^N k_j\label{mom}
\ee
where  $\l\{k_j\r\}_{j=1,\ldots,N}$ and $\l\{\la_\al\r\}_{\al=1,\ldots,M}$ satisfy a set of $N+M$ 
coupled Bethe equations, given below. 
The ground state momentum $p_0$ and the ground state energy $E_0$ are given uniquely 
by the location and the number of roots; especially, $p_0=0$. 
Starting from this ground state configuration, low-lying excitations are constructed by changing both the location and the number of the roots, thereby obtaining the corresponding momentum $p$ and energy $E$, both parametrized by the root sets. From these, one obtains the dispersion relation $E(p)$ and the excitation energy $\varepsilon(p)=E(p)-E_0$, which is analyzed according to Eq.~\refeq{disp}. The essential point is that the roots can be obtained explicitly in the limits of weak and strong coupling. The thermodynamic limit, which is essential for Eq.~\refeq{disp} to be defined at all, is carried out at the very end.  

\subsection{Derivation of results}
The Hamiltonian   
\begin{equation}
{H}=-\frac{\hbar ^2}{2m}\sum_{i = 1}^{N}\frac{\partial
^2}{\partial x_i^2}+\,g \sum_{1\leq i<j\leq N} \delta
(x_i-x_j)
\label{Ham-1}
\end{equation}
describes a quantum gas consisting of $N$ particles with equal masses in one dimension, where the particles interact via a pairwise $\delta$-function potential. According to the symmetry of the wave function, these may be bosons, fermions or a mixture of both. Here, we will consider a two-component Bose gas \cite{The-SB2} constrained by periodic boundary
conditions to a line of length $L$. The interaction is attractive for $c<0$ and repulsive for $c>0$.
The mass is denoted by $m$ and $g$ is the coupling constant which is written in terms of 
the scattering strength $c={2}/{a_{\rm 1D}}$ as $g ={\hbar ^2 c}/{m}$.
An effective $1$D scattering length $a_{\rm 1D}$ can be expressed through the 
$3$D scattering length for the bosons confined in a quasi $1$D geometry. 
For convenience in calculations, we prefer to set $\hbar=2m=1$, and thus to work with the parameter $c$. 
However, units may be restored at any time by using the dimensionless coupling constant 
\be
\gamma={mg}/{(\hbar^2n})={c}/{n}\label{gamdef}.
\ee 
The energy is given by Eq.~\refeq{en}, where 
the quasimomenta $k_j$ satisfy the Bethe equations (BE) \cite{The-SB2,GU,OBBG}
\begin{eqnarray}
& &\exp(\mathrm{i}k_jL)=-\prod^N_{\ell = 1} 
\frac{k_j-k_\ell+\mathrm{i}\, c}{k_j-k_\ell-\mathrm{i}\, c}\prod^M_{\mu = 1}
\frac{k_j-\lambda_\mu-\frac12 \mathrm{i} c}{k_j-\lambda_\mu+\frac12 \mathrm{i} c}\nonumber\\
& &\prod^N_{\ell = 1}
\frac{\lambda_{\alpha}-k_{\ell}-\frac12 \mathrm{i} c}{\lambda_{\alpha}-k_{\ell}+\frac12 \mathrm{i} c}
 = - {\prod^M_{ \beta = 1} }
\frac{\lambda_{\alpha}-\lambda_{\beta} -\mathrm{i}\, c}{\lambda_{\alpha}-\lambda_{\beta} +\mathrm{i}\, c} .
\label{BE}
\end{eqnarray}
Here $j = 1,\ldots, N$ and $\alpha = 1,\ldots, M$, with $M$ the number of spin-down bosons.

The ground state is the fully polarized state (total spin $S= N/2$) with $2 S+1$-fold degeneracy due to 
$su(2)$ symmetry \cite{The-SB1,OBBG}. This is achieved by choosing $M=0$ in Eq.~(\ref{BE}). 
So in the ground state, the BE reduce to those of the Lieb-Liniger Bose gas, 
which have been extensively studied via various methods \cite{KOR,Tbook,Wadati,BGM,Ger}.  
{}From previous work \cite{LL,Wadati,BGM,Ger} on the Lieb-Liniger Bose gas,
the ground state energy for weak coupling, namely $Lc \ll 1$, 
is given by 
\be
E_0\approx {N(N-1)c}/{L}\label{a}.
\ee   
In the strong coupling limit, i.e., as the strength of the repulsive interaction tends to infinity,
the bosons behave like impenetrable fermions and the system is known
as the Tonks-Girardeau gas. 
In this regime, the
ground state energy $E_0\approx \frac{\pi^2}{3L^2}\left(N^2-1\right)\left(1-{4n}/{c}\right)$
follows directly from the asymptotic solutions to the BE \cite{BGM}. 
Due to conformal invariance, the finite-size corrections to the ground state energy  
in the thermodynamic limit for strong coupling are given by
\begin{equation}
E_0(L,N)-Le^{(\infty)}_0=-\frac{ \pi C v_c}{6L} +o(1/L^2) \label{finite-size}
\end{equation} 
where $v_c=v_F\left(1-{4n}/{c}\right)$ with the Fermi velocity $v_F=2\pi n$ and the 
central charge $C=1$.
We notice that in the zero coupling regime $c\to 0$ the finite-size
corrections for the ground state energy do not meet the relation (\ref{finite-size}), because $v_c=2\sqrt{ c n}$ \cite{LL} (cf. also the discussion after Eq.~\refeq{BA-W}). 
This is due to the fact that for $c=0$, the Bose gas has dispersion $\hbar^2 k^2/(2 m)$, 
as can be seen from the Hamiltonian (\ref{Ham-1}). 
In that case the model is not conformally invariant (conformal invariance requires a {\em linear} dispersion). 
However, for any finite $c$, one identifies a finite Fermi momentum $k_F$. 
Then the dispersion relation can again be linearized around $\pm k_F$, yielding an effective 
bosonic theory which describes the density fluctuations with wave vectors $k\sim \pm k_F$ \cite{Haldane}. 
This theory is conformally invariant due to the linearized spectrum.

Spin charge separation is a typical phenomenon in interacting many-body systems. 
We first concentrate on the elementary excitations in the charge sector. 
These have linear dispersion, where the sound velocity is obtained from Eq.~\refeq{vmu}.
In terms of the Bethe roots, these excitations are created by particle-hole pairs with respect to the ground state configuration of the $\l\{k_j\r\}_{j=1,\ldots,M}$.
Let us first consider the weak-coupling limit, such that $k_i\propto \sqrt{c/L},\, i=1,\ldots, N$ \cite{BGM}.
Then, to order $c^2$, the equations
\begin{eqnarray}
\cos(k_iL)&\approx & 1-2\sum^{N}_{\ell=1}\frac{c^2}{(k_i-k_{\ell})^2}-4\sum_{\ell=1}^N\sum_{\ell < \ell '=2}^{N}\frac{c}{(k_i-k_{\ell})}\frac{c}{(k_i-k_{\ell '})}\label{BE-app1}\\
\sin(k_iL)& \approx & \sum_{\ell=1}^N\frac{2c}{(k_i-k_{\ell})} 
\label{BE-app2}
\end{eqnarray}
hold, which determine the roots of the BE (\ref{BE}) asymptotically. 
The lowest possible excitations are obtained from the set of equations \cite{BGM}
\begin{eqnarray}
k_N&\approx& p+\frac{2c}{L}\sum_{\ell=1}^{N-1}\frac{1}{k_N-k_\ell}\nonumber\\
k_j &\approx&\frac{2c}{L}\sum_{\ell=1,\ell\neq j}^{N}\frac{1}{k_j-k_\ell} 
\label{BA-W}
\end{eqnarray}
where $j=1,\ldots,N-1$ and the momentum $p$ parametrizes the
excitation. Here, $p=2 \ell \pi/L$ with $\ell$ integer; however, we will
treat $p$ as a variable in the following, which will take continuous values in the thermodynamic limit. Note that $p$ is introduced such that Eq.~\refeq{mom} holds.  
The excited energy in the long wavelength limit is easily calculated from
equations (\ref{BA-W}), i.e., $E(p)\approx N(N-1)c/L+pk_B+p^2$,
where the Fermi boundary $k_B\approx 2\sqrt{cn}$ is obtained from the Hermite
polynomial description \cite{Fermi-Boundary}.
%
%
%
Therefore $E(p)-E_0 \approx  2\sqrt{cn} p= v_cp$. 
Here the charge velocity is given by $v_c=2\sqrt{nc}$, which coincides with the result obtained from \refeq{a} according to the formula
$v_c=\sqrt{\frac{2L}{n}\left[\frac{\partial ^2 E}{\partial L^2}\right]}$ \cite{LL}.  
{}From the Bethe roots (\ref{BA-W}), the collective effect is that if one particle jumps out of the
Dirac sea the rest also rearranges.

Similarly, for the strong coupling limit $Lc\gg 1$, if the largest
quasimomentum $k_N$ jumps out of the Dirac sea, asymptotic
roots of the BE (\ref{BE}) can be obtained from a $1/(Lc)$ expansion as
\begin{eqnarray}
k_N&\approx& \left(\frac{(N-1)\pi}{L}+p+\frac{2p}{Lc}
\right)\left(1-\frac{2N}{Lc}\right)\nonumber\\
k_j&\approx & \left(\frac{2n_j\pi}{L}+\frac{2p}{Lc} \right)\left(1-\frac{2N}{Lc}\right)
\end{eqnarray}
where $n_j=\pm 1, \pm 3, \ldots, \pm (N-3)/2, -(N-1)/2$ if $N$ is odd. 
It follows that $E(p)-E_0\approx
\frac{2(N-1)\pi}{L}\left(1-\frac{2N}{Lc}\right)^2p\approx  v_cp$ with
the charge velocity $v_c= \frac{2N\pi}{L}\left(1-\frac{4N}{Lc}\right)$
and the ground state energy $E_0\approx \frac{\pi^2}{3L^2}N(N^2-1)(1-\frac{4N}{Lc})$. 
Note that as $c\to \infty$ the velocity $v_c$ becomes the Fermi velocity of noninteracting spinless fermions. 
%


We now focus on the spin excitation with regard to one spin flipping. 
This corresponds to introducing one $\lambda$-solution into Eqs.~\refeq{BE}, which we call $\lambda_1$.
In the long wavelength limit a quadratic dispersion of spin wave excitations above the 
ferromagnetic ground state is obtained \cite{SW-1,SW-2,The-SB3}
\begin{equation}
E(p)-E_0 \approx p^2/2m^{*} \label{eff-mass}
\end{equation}   
where $m^{*}$ is an effective mass. This is the case indicated in Eq.~\refeq{disp-2}.
Now the small-coupling expansion of the BE (\ref{BE}) yields
\begin{equation}
k_j\approx
\frac{2c}{L}\sum_{\ell=1}^N\frac{1}{k_j-k_\ell}+\frac{c}{L\lambda_1}+\frac{c}{L\lambda_1^2}k_j
\label{spin-w}
\end{equation}
for $j=1,\ldots, N$. From Eq.~\refeq{mom}, one obtains $\lambda _1 \approx \frac{Nc}{Lp}$, which in the long-wavelength limit diverges as $1/p$, so that $\lambda_1\gg k_B$. We now calculate the energy by inserting Eq.~\refeq{spin-w} into Eq.~\refeq{en},
\begin{eqnarray}
E(p)&=& \sum_{j=1}^N k_j^2\approx 
\left[\frac{c}{L}N(N-1)+\frac{c}{L\lambda _1}p\right]
\l(1+\frac{c}{L\lambda_1^2}\r) = E_0+p^2 \label{S-W}
\end{eqnarray}
which indicates ${m}/{m^{*}}\approx 1$,  where $m$ is the boson mass.
In  the above equation $E_0$ is given by Eq.~\refeq{a}.
We notice that higher order corrections to the ground state energy are very 
hard to derive from our expansion ansatz.  
The next correction to the ground state energy of the Lieb-Liniger Bose gas is 
known to be given by 
$E_0\approx \frac{N^2c}{L} \left(1-\frac{4\sqrt{c/n}}{3\pi}\right)$ \cite{Wadati}.  
Thus from the relation (\ref{S-W}) we observe that
\begin{equation}
E(p)\approx E_0\left(1+\frac{c}{L\lambda_1^2}\right).
\end{equation}
This suggests the effective mass ${m}/{m^{*}}\approx 1-\frac{4\sqrt{c/n}}{3\pi}$.


The divergence $\lambda_1\sim 1/p$ continues to hold in the strong coupling regime $c \gg 1$.
In this regime, the  consistency conditions derived from the BE
(\ref{BE}) in the $1/(Lc)$ expansion are 
\begin{equation}
\sin k_jL\approx -2\sum_{\ell=1}^N\frac{k_j-k_\ell}{c}+\frac{c}{\lambda_1}\left(1+\frac{k_j}{\lambda_1}\right).
\end{equation}
Here for convenience we consider $N$ as an odd number.  
Then  
\begin{equation}
k_j\approx \left(\frac{2\pi n_j}{L}+\frac{2p}{Lc}+\frac{c}{L\lambda_1}\right)\left(1-\frac{2N}{Lc}+\frac{c}{L\lambda_1^2}\right)
\end{equation}
where $n_j=\pm 1, \pm 3,\ldots, \pm (N-1)/2$. 
{}From Eq.~\refeq{mom}, 
we find $\lambda_1\approx Nc/Lp \gg c \gg k_j$. 
After some algebra we obtain
\begin{eqnarray}
E(p)& =& \sum_{j=1}^Nk_j^2 \approx
\frac{\pi^2}{3L^2}N(N^2-1)\left(1-\frac{4N}{Lc}\right)\nonumber\\
& &+p^2\left(\frac{1}{N}+\frac{2\pi^2N}{3Lc}\left(1+\frac{2N}{Lc} \right)+\frac{4}{Lc}\right)\left(1-\frac{4N}{Lc}\right)
\end{eqnarray} 
which suggests the effective mass
\begin{equation}
\frac{m}{m^*}\approx \frac{1}{N}+\frac{2\pi^2n}{3c}\left(1-\frac{2n}{c}\right).
\end{equation}
The leading correction term coincides with the earlier result \cite{The-SB3}. 
The  next correction term is relevant for finite $c$.  
In the strong coupling limit, i.e., $c\to \infty$, all the bosons become
indistinguishable thus behaving like hardcore bosons with Fermi pressure-like kinetic energy. 
The effective mass takes the maximum value $m^*=Nm$, meaning that by
moving one boson with down spin, one has to move all the particles
with up spins.  
In this way the spin velocity is very small, as if the spin-flip is frozen \cite{The-SB3}.


\section{The Fermi gas}

The spin-$\frac12$ Fermi gas with equal number of spin-up and spin-down
fermions described by the Hamiltonian (\ref{Ham-1}) has been extensively studied 
in the context of the BCS-BEC crossover
\cite{BEC-BCS1,BEC-BCS2,BEC-BCS3,BEC-BCS4}.  
Very recently, a comprehensive study of the model with arbitrary polarization 
was undertaken \cite{BBGO-F}. 
In order to compare different collective dynamics of elementary excitations in the
charge and spin degrees with those of the spinor Bose gas, we derive the 
charge and spin velocities for the strongly attractive  Fermi gas. 
This is done from an expansion of the discrete BE, similarly to the previous section. 
At the end of this section, we will compare the results with those obtained from the 
BE in the thermodynamic limit.

\subsection{Velocities from the discrete Bethe ansatz equations}

In the strongly attractive regime the bound states behave like bosonic 
tightly bound molecular dimers. 
The ground state consists of two-body bound states spread out about the origin in
pseudomomentum space due to Fermi pressure,
i.e., with $k_j=\la_j+\frac12 \mathrm{i}c,\,k_{-j}=\la_j-\frac12\mathrm{i}c$
for $j=1,\ldots,N/2$, where the parameters $\left\{\la_j\right\}_{j=1,\ldots,M}$
satisfy the BE \cite{GY,Tbook}
\begin{equation}
\te^{\mathrm{i}\,2\la_j L}=-\prod_{\ell=1}^M
\frac{\la _j-\la_\ell+\mathrm{i}c}{\la _j-\la_\ell-\mathrm{i}c},\qquad M={N}/{2}
\label{BE-F}
\end{equation}
for $j=1,\ldots,M$. 
The lowest  charge excitation comes from the configuration in which  a bound state
  with the largest momentum $k_{\rm
  max}=\frac{(M-1)\pi}{L}(1-\frac{M}{Lc})$, which is at the edge of
  the Dirac sea in quasimomentum space, jumps out of the Dirac sea. 
This leads to a bound  state density fluctuation. 
The total real momentum changes from zero to $p$.  
Taking $M$ even, we may obtain the roots of the BE (\ref{BE-F}) as
\begin{eqnarray}
\la _M&\approx &
\left[\frac{(M-1)\pi}{2L}+\frac{p}{2}+\frac{p}{2Lc}\right]\left(1-\frac{M}{Lc}\right)\nonumber\\
\la_j&\approx &\left[\frac{2n_j-1}{2L}+\frac{p}{2Lc} \right]\left(1-\frac{M}{Lc}\right)
\end{eqnarray}
where $j=1,\ldots, M-1$ and $n_j=-(M-2)/2,-(M-4)/2,\ldots, (M-2)/2$. 
It follows that
\begin{eqnarray}
\fl \qquad
E(p)\approx
-\frac14 {c^2} N +\frac{\pi^2}{48L^2}N(N^2-2)\left(1-\frac{2M}{Lc}\right) 
+\frac{(M-1)\pi}{L}\left(1-\frac{2M}{Lc}\right)p.
\end{eqnarray}
It is seen that to this order $E(p)-E_0 = v_c\,p$ with charge velocity 
\be
v_c=\frac{n\pi}{2}\left(1-\frac{n}{c}\right)
\label{vcapprox}
\ee
which coincides
with the result of Ref. \cite{BEC-BCS1} and with the result derived from the ground state energy $E_0$
\cite{BEC-BCS1,BEC-BCS2,BEC-BCS4,BBGO-F},
\begin{equation}
E_0=-\frac14 {c^2} N +\frac{\pi^2}{48L^2}N(N^2-2)\left(1-\frac{2M}{Lc}\right).\label{F-GE}
\end{equation}
Again we may find from the ground state energy that the finite-size
corrections in the strongly repulsive limit for the Fermi gas also meet the relation
(\ref{finite-size}) with central charge $C=1$. The spin-channel does not contribute here, 
because it is gapped (cf. below).

We now focus on the spin velocity which involves breaking one bound
state over the ground state in order to get spin flipping. 
The ground state is a spin singlet. 
The lowest spin excitation is incurred by a spin triplet excitation with total spin $M^z=1$. 
In such a configuration the BE become \cite{Tbook}
\begin{eqnarray}
&&\te^{\mathrm{i}\,2\la_j L}=-\prod_{\ell=1}^{M-1}\frac{\la
  _j-\la_\ell+\mathrm{i}c}{\la
  _j-\la_\ell-\mathrm{i}c}\prod_{\mu=1}^{2}\frac{\la
  _j-k_\mu+\frac12 \mathrm{i} c}{\la _j-k_\mu-\frac12 \mathrm{i} c}\nonumber\\
&&
\te^{\mathrm{i}\,k_\mu L}=-\prod_{\ell=1}^{M-1}\frac{k_\mu
  -\la_\ell+\frac12 \mathrm{i} c}{k_\mu
 -\la_\ell-\frac12 \mathrm{i} c} \label{F-BA}
\end{eqnarray} 
with $j=1,\ldots M-1$ and $\mu=1,2$. 
Withought loss of generality, we assume $M$ is even. 
The BE (\ref{F-BA}) admit  the asymptotic roots
\begin{eqnarray}
k_1 &\approx &\left(\frac{n_{+}\pi}{L}+\frac{4}{Lc}\sum_{j=1}^{M-1}\la _j  \right)
\left(1-\frac{4(M-1)}{Lc}\right)\nonumber\\ 
k_2 &\approx &\left(\frac{n_{-}\pi}{L}+\frac{4}{Lc}\sum_{j=1}^{M-1}\la _j  \right)
\left(1-\frac{4(M-1)}{Lc}\right)\nonumber\\ 
\la _j &\approx & 
\left(\frac{n_j\pi}{L}+\frac{1}{Lc}\sum_{j=1}^{M-1}
\la_j+\frac{2(k_1+k_2)}{Lc}\right)\left(1-\frac{M-3}{Lc}\right).
\label{F-BA-2}
\end{eqnarray}
In the above equations $n_j=0,\pm 1,\cdots, \pm (M-2)/2$ and $n_{\pm
  } $ are odd numbers to be determined.
The number of bound states is now $M-1$. 
This leads to a different parity for the BE (\ref{F-BA}). 
We can see this directly from the roots (\ref{F-BA-2}) where all bound states shift by nearly $\pi/2$
from their original  positions in the ground state configuration. 
The unpaired two fermions are scattered away.
Summing up both sides of the third equation in Eq.~(\ref{F-BA-2}) gives the relation 
$\sum_{j=1}^{M-1}\la_j\approx \frac{2(M-1)}{Lc}(k_1+k_2)$. 
This relation implies that $\sum_{j=1}^{M-1}\la_j $ is negligible in comparison with
$k_1+k_2$ in the strong coupling regime.  
Further, from Eq.~\refeq{mom} we have 
\begin{equation}
k_1+k_2\approx p\left(1-\frac{4(M-1)}{Lc}\right)
\end{equation}
if we consider  $n_{-}=-n_{+}$ for minimizing the energy.
Using the BE in (\ref{F-BA}) gives another  relation 
\begin{equation}
k_1-k_2\approx
\frac{2n_{+}\pi }{L}\left(1-\frac{4(M-1)}{Lc}\right).
\end{equation}
Thus we find all roots to be given by
\begin{eqnarray}
\la_j&\approx & \left(\frac{n_j\pi}{L}+\frac{2p}{Lc}(1-\frac{4(M-1)}{Lc})\right)
\left(1-\frac{M-3}{Lc}\right)\nonumber\\
k_{1,2}&\approx &\left(\pm \frac{n_{+}\pi}{L}+\frac{p}{2} \right)\left( 1-\frac{4(M-1)}{Lc}\right).
\end{eqnarray}
Here $n_j=0,\pm 1,\ldots, \pm (M-2)/2$. 
The quantum numbers $n_{\pm }$ for the unpaired fermions should be chosen to be 
as small as possible, i.e., $\pm 1$, in order to minimize the energy \cite{Tbook}. 
Actually, $n_\pm$ may take any two odd integer values, which correspond to states with 
energy higher than the ground state energy. 
After some algebra, we obtain the excitation  energy
\begin{eqnarray}
E(p)&\approx &
-2(M-1)\frac{c^2}{4}+\frac{\pi^2}{48L^2}\left(1-\frac{2(M-3)}{Lc}\right)N(N^2-4)\nonumber\\
& & +\frac12\,{p^2}\left(1-\frac{8(M-1)}{Lc}\right) +O\l(\frac{n_+^2}{L^2}, \frac{M^2}{L^2}\r) \label{ep}.
\end{eqnarray}
The choice of $n_\pm$ affects the excitation energy only to the indicated order. 
{}From Eq.~\refeq{ep} we deduce the dispersion relation 
$\varepsilon(p)=E(p)-E_0= \Delta+\frac12\,{p^2}\left(1-\frac{8M}{Lc}\right)$, where the
spin excitation gap to leading order in $c$ is $\Delta \approx c^2/2$ and $E_0$ is given by (\ref{F-GE}). 
On the other hand, by comparing with Eq.~(\ref{disp}), the spin velocity follows as
\begin{equation}
v_s = \sqrt{\Delta}\left(1-\frac{4M}{Lc}\right)=
\frac{|c|}{\sqrt{2}}\left(1-\frac{2n}{c}\right)\label{vsapprox}
\end{equation}
which is divergent due to the gapped spin triplet excitation
\cite{SC,BBGO-F,Cazalilla}. This result has been obtained previously in Ref. \cite{BEC-BCS1}.

\subsection{Velocities from the Bethe ansatz equations in the thermodynamic limit}

\begin{figure}[h]
\begin{center}
\vspace{1cm}
\includegraphics[scale=0.5]{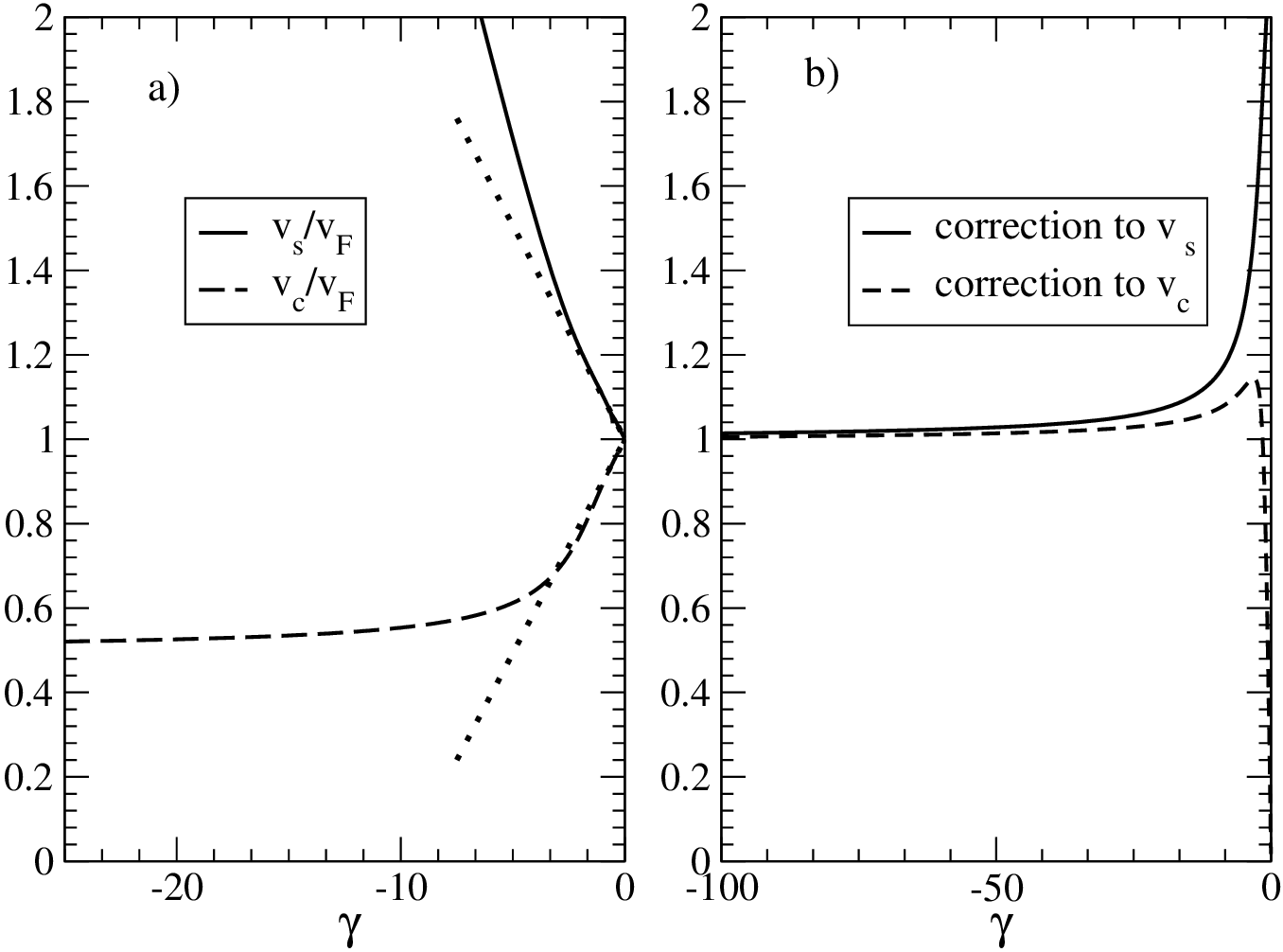}
\caption{a) Charge and spin velocity divided by $v_F=\pi n$ for the
  attractive regime, dependent on $\g$. The dashed lines are the small
  $\g$ approximations, $v_{c,s}/v_F=1\pm \g/\pi^2$. b) The corrections
  to $v_{c,s}$ in the strongly attractive limit, namely
  $-\g(2v_c/v_F-1)$ and $(\pi \sqrt2 v_s/(v_F \g)+1)\g/2$. According
  to Eqs.~\refeq{vcapprox}, \refeq{vsapprox} and \refeq{gamdef}, these expressions approach 1 in the limit $\g\to -\infty$, which is confirmed numerically.}  
\label{fig1}
\end{center}
\end{figure}

Spin and charge velocites for the spin-$\frac12$ Fermi gas in the
attractive regime have been calculated  \cite{BEC-BCS1,BBGO-F}, using
the BE in the thermodynamic limit. Within this approach, the ideas sketched in
Section \ref{sec21} are applied to the log form of
Eqs.~\refeq{BE}, thereby converting products to sums. By introducing
densities, these can in turn be converted to integrals. One ends up
with linear integral equations which can be solved numerically in
order to determine $v_c$ and $v_s$. 
The advantage of this approach is that it is valid for arbitrary couplings. 
In Fig.~\ref{fig1}, $v_c$ and $v_s$ are depicted from a numerical solution of 
those integral equations for attractive fermions. Furthermore, 
the results \refeq{vcapprox}, \refeq{vsapprox}, are confirmed.

\section{Concluding remarks}

In conclusion, we have derived explicitly the dispersion relations for elementary
charge and spin excitations for the two-component Bose and Fermi
gases in the long wavelength limit.
Analytic results have been found for the charge and spin velocities and for the 
effective mass connected with spin wave excitations in the repulsive spinor Bose gas.  
These quantities reveal significant insight into the nature of the
collective hydrodynamics of the Bose and Fermi gases.
In particular, physical quantities such as the charge velocity, spin
velocity, effective mass, Fermi momentum, finite-size corrections,
ground state and excitation energies, are obtained via an analysis of the discrete Bethe ansatz solutions, as opposed to using
the continuous integral equation approach.
As such we believe that it can be easily adapted to a number of related models.

\ack
This work has been supported by the Australian Research Council and the 
German Science Foundation under grant number BO 2538/1-1. 
We thank D. M. Gangardt, C. Lee, S. Sergeev  and H.-Q. Zhou for helpful discussions.  
XWG also thanks the Department of Mathematics, University of Science and Technology of
China, for kind hospitality.

\end{document}